**Library Philosophy and Practice (e-journal)**  Libraries at University of Nebraska-Lincoln

# Research Data Management and Services in South Asian Academic Libraries

Jahnavi Yidavalapati  
*National Institute of Agricultural Extension Management, Hyderabad*, jahnavi.sneha@gmail.com

Priyanka Sinha  
*Punjab University*, priyankasinha101099@gmail.com

Subaveerapandiyan A  
*Regional Institute of Education Mysore*, subaveerapandiyan@gmail.com

# Research Data Management and Services in South Asian Academic Libraries

**Jahnavi Yidavalapati, Priyanka Sinha and Subaveerapandiyan  A .**
Research Associate (Documentation), National Institute of Agricultural Extension Management, Hyderabad, Telangana
PhD. Research Scholar, Department of Library and Information Science, Punjab University
Professional Assistant (Library), Regional Institute of Education Mysore, India
Email: subaveerapandiyan@gmail.com

**Abstract**
The study examined the research data management and related services offered by South Asian countries' academic libraries. Research applied quantitative approach and survey research design method were used for this study. The survey questionnaire was distributed randomly to academic library professionals in five countries: Afghanistan, Bangladesh, India, Pakistan, and Sri Lanka. The sample population comprised 67 library professionals from various institutes of five countries. The study's findings revealed that 83.6% of institute libraries provide research data management services to their users. The study recommends that institutes or funding organizations support staff to attend conferences and workshops on research data management, library professionals have to join MOOC to take courses related to research data services, Institute or professionals conduct in-house staff workshops and presentations. The study also found that 64.2% agreed compliance with funder requirements and preservation are major issues.

1. Introduction

Many academic libraries have adopted research data management (RDM) services to support research activity at the university level. Academic libraries act as repositories of research works, dissertations created by various research scholars and students. In this information era, all works are created and stored digitally in a massive quantity of data and information that makes it difficult to find authentic information resources. To overcome this, research data management services are adopted. The primary goal of research data management services is to gain maximum data from their investigation or research project (Chiware & Mathe 2016).

### 1.1 What is RDM?
Research data management describes management of data in an entire research project or work. It is a process of creating or collecting the data, storing, organizing, and maintaining the data. This service supports data management planning, digital and metadata curation and include conversion (Tenopir, Sandusky & Birch, 2014).

"RDM consists of several different activities and processes associated with the data lifecycle, involving the design and creation of data, storage, security, preservation, retrieval, sharing, and reuse, all taking into account technical capabilities, ethical considerations, legal issues, and governance frameworks". (Cox et.al 2014).

RDM deals with Data documentation (metadata), Data organisation (file formats, data exchange), Data storage and archiving, Legal aspects (copyright, data protection, licensing), Data publication (repository, data centres, persistent identifier (PID)).

The library staff has always been supportive of research scholars and students. With the changing environment, library staff are also upgrading their services. This study collected the data through a survey method showing the present RDM services scenario in South Asian academic libraries. The results will inform about the issues and challenges faced by libraries to deliver effective RDM services and highlight librarian staff's approaches in skill development.

2. **Objectives of the study**
   - ➢ To identify South Asian academic libraries are providing RDM services.
   - ➢ To evaluate how frequently RDM services are performed by the libraries.
   - ➢ To check RDM current services offered by libraries.
   - ➢ To know skills and competencies required for providing RDM services.
   - ➢ To examine RDM skill development needs among library professionals.
   - ➢ To understand the RDM management challenges and issues.

3. **Review of Literature**

Tenopir et al. (2014), investigated research data services offered by two countries' research libraries in the USA and Canada. Research data services (RDS) are divided into two categories: informational services and technical services. In their survey study they found out still RDS services are in the planning stage. 49.5% of libraries only gave the find data set and reference services remaining are planned to implement within two years and similarly most of libraries still did not plan for RDS services in the library. They found informational services were more common than technical services in library data services. They also identified challenges and issues facing academic libraries concerning RDS, such as limited data management skills of library staff and insufficient resources assigned to data-related services. Finally, the researcher suggests that RDS services skill related workshops and training needs to the librarians also.

Gordon et al. (2015) discusses the Databrary repository working model and how it supports the researchers in their research activity. The study revealed that the participants perceived data archiving, data preservation, and data documentation as more critical in RDS than other types of services as libraries have a suite of expertise in archiving and preservation. In his opinion, a sound library team under good leadership can make a good repository that helps in boosting the RDS service.

Kennan, M.A. (2016) investigated the knowledge and skill requirements for data management in universities and scientific research organizations. The author interviewed 25 data professionals in universities and scientific research organizations. The author describes skills and knowledge that are required like interpersonal skills, behavioral skills, ICT skills and legal and regulatory frameworks in this digital era. He thinks contextual knowledge is necessary for understanding and highlighted the need to develop effective communication.

Yu et al. (2017) describes the implementation of RDM services in the University of Queensland Library through the Research Life cycle model. This model is categorized into three categories: 1. the project planning and preparing stage. 2. The project has conducted the stage. 3. The archiving, publishing and disseminating stage and details the strategies in designing and delivering RDM services, which include preparing guides, designing training programs for faculty, librarians, researchers and engaging stakeholders within and external to the university. The author mentioned that due to deliberate leadership effect and support from the staff, the University of Queensland Library delivers effective RDM services to researchers.

Cox et al. (2017), investigated developments of Research Data Management in academic libraries in seven countries (Australia, Canada, Germany, Ireland, Netherlands, New Zealand and UK). The author focuses on development in the areas of RDM policy and governance, service developments, and staff deployment and skills. In their RDM maturity model survey, they specified four levels of RDM maturity i.e., Level 0 as "none," Level 1 as "basic," Level 2 as "developing," and Level 3 as "extensive". The author survey reveals that the RDM services reached "Extensive". Even with this RDM maturity model, substantial empirical evidence is needed to test, verify, or enhance the model. He thinks the libraries should give more priority to upgrading RDM services.

Shelly & Jackson (2018) describes the role of libraries to support RDM services. The aim of the study was to identify university groups and the role of libraries to provide RDM services. The study was based on primary data and to collect information. 13 Australian universities were examined using the content analysis method. Generally, strong encouragement was given to secure and store research data during and after the project. The article concluded that libraries were quite active to support the RDM services. There was a need for advice and practical suggestions to researchers on RDM, particularly in the areas of creating metadata and loading data to repositories.

Zhou (2018) examined the perceptions and practices of Academic libraries to provide RDM service. This paper aims to explore RDM services and effective recommendations for academic libraries to conduct data management services. The study identified many core elements of RDM service practices such as policy design, architecture, service quality, funding model, and staffing. The author concludes that the University RDM service is still in its infancy and recommends Universities and their libraries need to have a deep understanding of the above-mentioned factors.

Frederick & Run. (2019) examined 81 Ghanaian University libraries to explore the role of academic libraries in research data management. The authors gathered data using a semi-structured interview method. The study reveals that university libraries are at the beginning stage to provide Research data service to their researchers. Some libraries have completed the policy framework while some are being formulated. The author recommends the libraries to create an RDM community to provide help in developing the skills of staff, establishing a data repository and involve in campaigns held with academic staff and researchers.

Gowen & Meier (2020) examined libraries at the 60 U.S. AAU Institution research services for the past 5 years. This study reveals that the support of research data management and services are becoming more prominent, but libraries are failing to maintain the positions of Data librarians. The authors recommend that the libraries should rethink upgrading their RDM services.

## 4. Materials and Methods

The study is exploratory research design, quantitative approach, survey method, simple random sampling, questionnaire format, tool used for this study was Google Form link shared by mail. The research data was collected from South Asian Library and Information Science professionals who are those working in Librarian to Library Assistant. The present study samples belong to Afghanistan, Bangladesh, India, Pakistan, Sri Lanka and excluded Bhutan, Nepal and Maldives. The questionnaire consists of two parts. Part one demographic one question country details. Part two consists of three questions related to RDM and RDS services. Part three consists of three questions related to RDM skills development needs, Skills and competencies required for RDM services, Approaches of approaches to staff RDM skills development in libraries. Part 4 consists of one question which was RDM managerial issues and challenges. Data analysis and diagram making tool used Google Excel Sheet. Data analysis simple frequency and percentage.

## 5. Data Analysis and Interpretation

**Table 1: Survey responses by country**

| Survey responses by country | Frequency | Percentage |
|---|---|---|
| Afghanistan | 1 | 1.5 |
| Bangladesh | 13 | 19.4 |
| India | 35 | 52.2 |
| Pakistan | 11 | 16.4 |
| Sri Lanka | 7 | 10.4 |

**Figure 1. Survey responses by country**

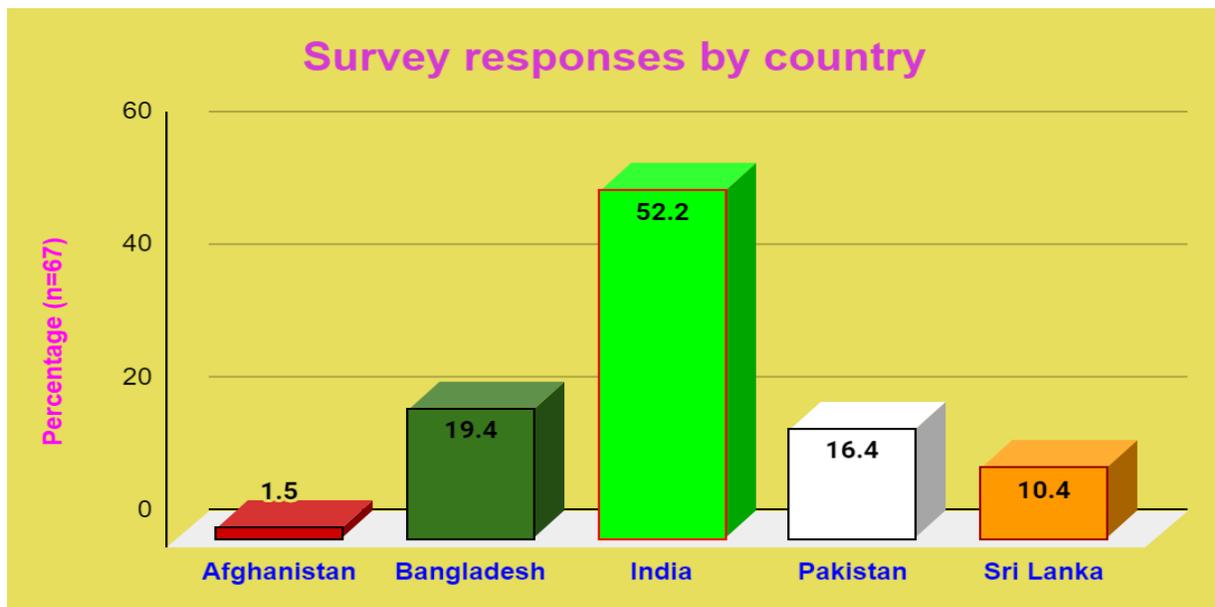

The table 1 and figure 1 indicate the respondents' country. The statistics show us that most of the respondents are from India 52.2%, and least respondents are from Africa 1.5%. The remaining countries are Bangladesh 19.4%, followed by Pakistan 16.4% and Sri Lanka 10.4% of respondents.

**Table 2. Provision of RDM services by institute**

| Does your institute provide RDM services? | Frequency | Percentage |
|---|---|---|
| Yes | 56 | 83.6 |
| No | 11 | 16.4 |

**Figure 2. Provision of RDM services by institute**

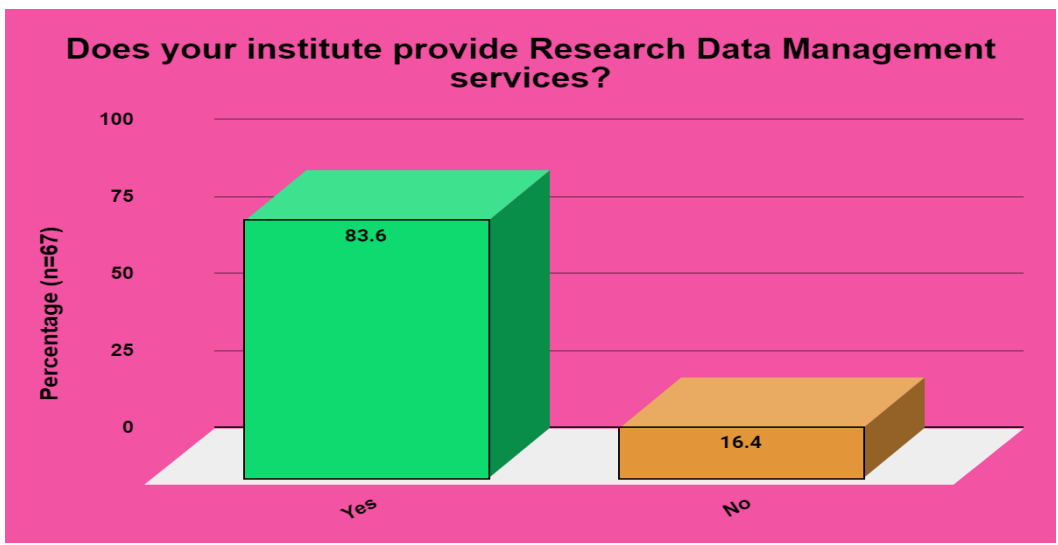

The table 2 and figure 2 indicate research data management services provided by the institutes. This result shows that 83.6% of institutes offering RDM services and 16.4% are not providing RDM services to the library users.

**Table 3. Frequency of performing each RDM services**

| How frequently performed each of the RDM | Frequency | Percentage |
|---|---|---|
| Never performed. | 11 | 16.4 |
| Performed a few times a year. | 20 | 29.9 |
| Performed about once a month | 11 | 16.4 |
| Performed about once a week | 14 | 20.9 |
| Performed daily | 11 | 16.4 |

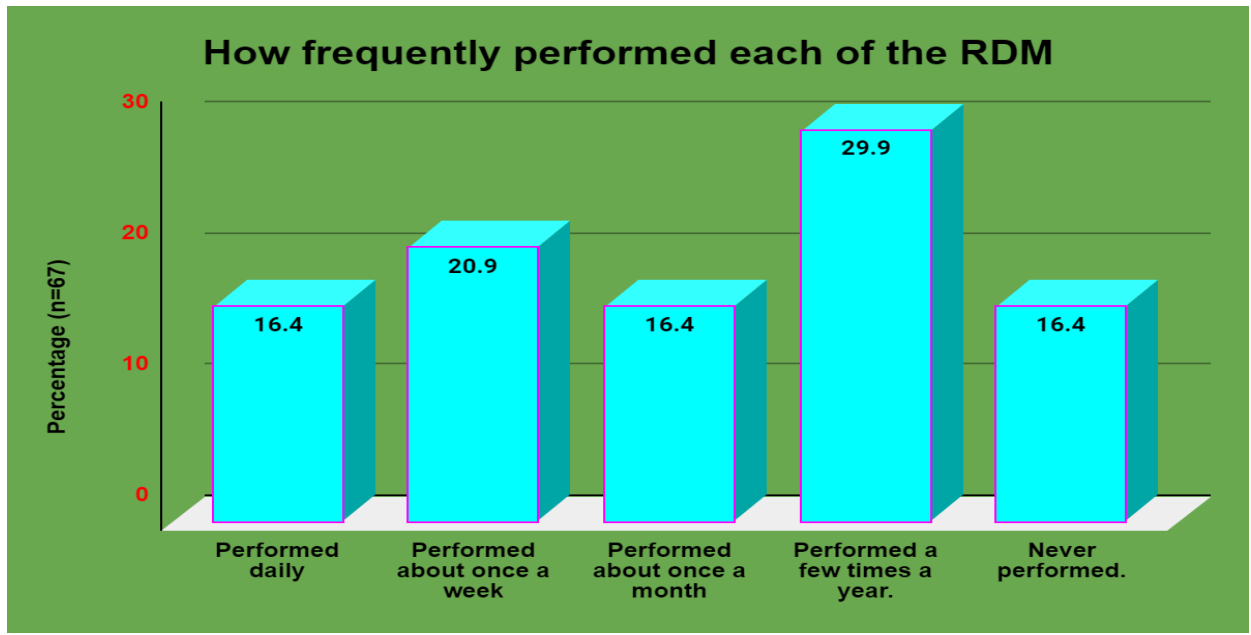

**Figure 3. Frequency of performing each RDM services**

Table 3 and figure 3 clearly shows the frequency of research data management services performed. The result of the analysis reveals 16.4% institutes performed daily and month, equally 16.4% never performed because that institutes are not offering the services so it does not perform. Most of the institutes 29.9% performed few times in a year, followed by 20.9% institutes performed once in a week.

**Table 4: RDM current services in libraries**

| RDM current services in libraries | Frequency | Percentage (n=67) |
|---|---|---|
| Analyses and visualize data sets using Python scripts, SPSS, R and MS Excel software | 22 | 32.8 |
| Carrying out long term preservation of research data (Data repositories/ Institutional Repository) | 26 | 38.8 |
| Data management planning (DMP) advisory service | 16 | 23.9 |
| Data management training and/or data literacy instruction (e.g., to research students, early career researchers, etc. | 27 | 40.3 |
| Data Publishing, Sharing & Reuse (Intellectual property assistance, Metadata assistance) | 14 | 20.9 |
| Data Study & Analysis (Instructional Support) | 22 | 32.8 |
| Maintaining a web resource/guide of local advice and useful resources for RDM | 19 | 28.4 |
| Offer an advisory service on data analysis/mining/visualization | 11 | 16.4 |

| Promote awareness of reusable data sources, such as data archives | 20 | 29.9 |
|---|---|---|
| Support reproducibility, transparency in workflows and research integrity | 15 | 22.4 |
| We did not provide any services | 11 | 16.4 |

**Figure 4. RDM current services in libraries**

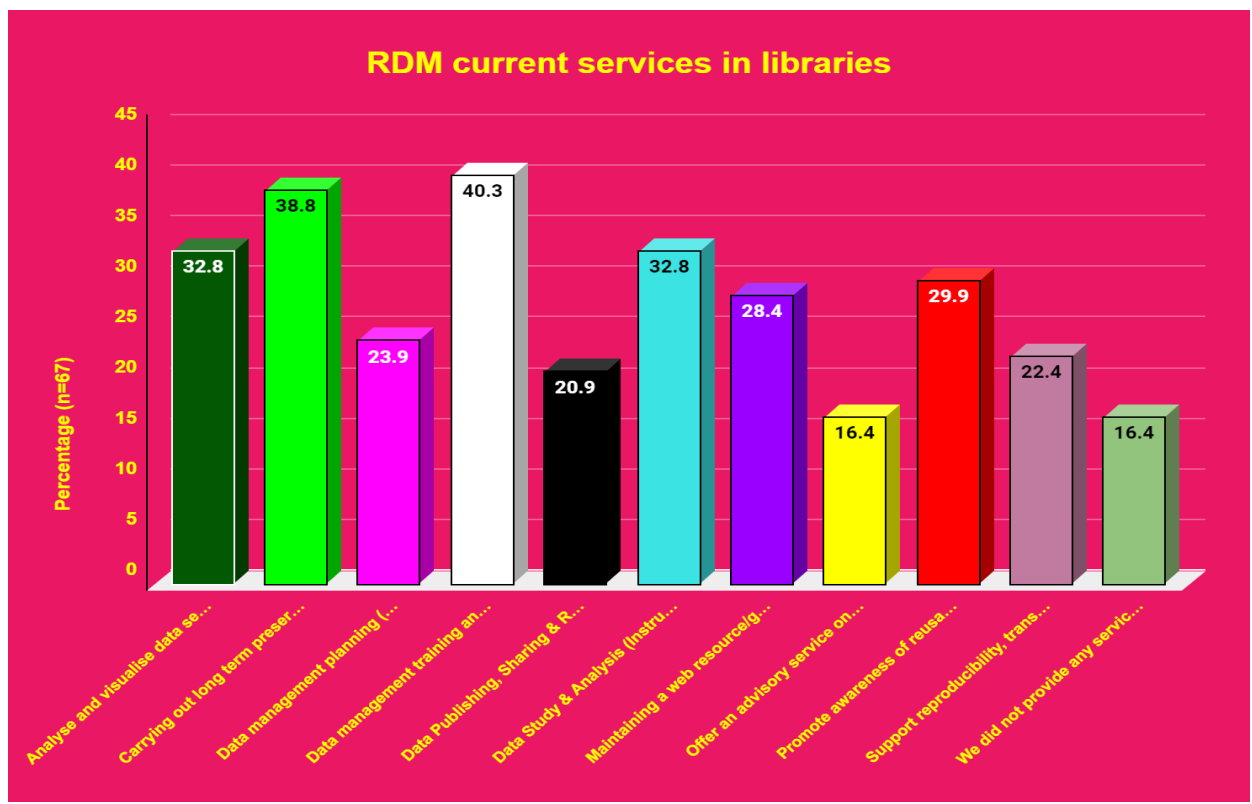

The above percentage analysis shows that (table and figure 4) research data management services offered by the libraries at present. Most of the institutes provide 40.3% data management training and/or data literacy instruction (e.g. to research students, early career researchers, etc., followed by many other services also provided by the libraries they are 38.8% carrying out long term preservation of research data such as data repositories and institutional repository services, 32.8% analyses and visualize data sets using Python scripts, SPSS, R and MS Excel etc.; data study & analysis instructional support services, 29.9% institute doing the awareness of reusable data sources, such as data archives. Similarly, 28.4% maintaining a web resource/guide of local advice and useful resources for RDM, followed by 23.9% data management planning and advisory services, 22.4% supports reproducibility, transparency in workflows and research integrity, 20.4% data publishing, sharing and reuse (intellectual property assistance, metadata assistance), 16.4% Offers an advisory service on data analysis/mining/visualization. 16.4% of institutes did not provide any services.

**Table 5. Library RDM skill development needs**

| Library RDM skill development needs | Strongly Agree | Agree | Disagree | Strongly Disagree |
|---|---|---|---|---|
| Data curation & Metadata skills | 24 (35.8%) | 40 (59.7%) | 3 (4.5%) | 0 (0%) |
| Data description and documentation | 20 (29.8%) | 45 (67.2%) | 2 (3%) | 0 (0%) |
| Knowledge of institutional and extra-institutional resources | 30 (44.7%) | 32 (47.8%) | 3 (4.5%) | 2 (3%) |
| Knowledge of RDM principles, relevant technologies and processes, metadata | 25 (37.3%) | 39 (58.2%) | 3 (4.5%) | 0 (0%) |
| Knowledge of researchers' needs, knowledge of available material | 37 (55.2%) | 27 (40.3%) | 3 (4.5%) | 0 (0%) |
| Knowledge of the research lifecycle | 24 (35.8%) | 36 (53.7%) | 6 (9%) | 1 (1.5%) |
| Legal, policy and advisory skills (e.g., intellectual property, ethics, licensing etc.) | 23 (34.3%) | 40 (59.7%) | 4 (6%) | 0 (0%) |
| Technical and ICT skills (e.g., data storage, infrastructure, architecture etc.) | 34 (50.7%) | 31 (46.3%) | 2 (3%) | 0 (0%) |

**Figure 5. Library RDM skill development needs**

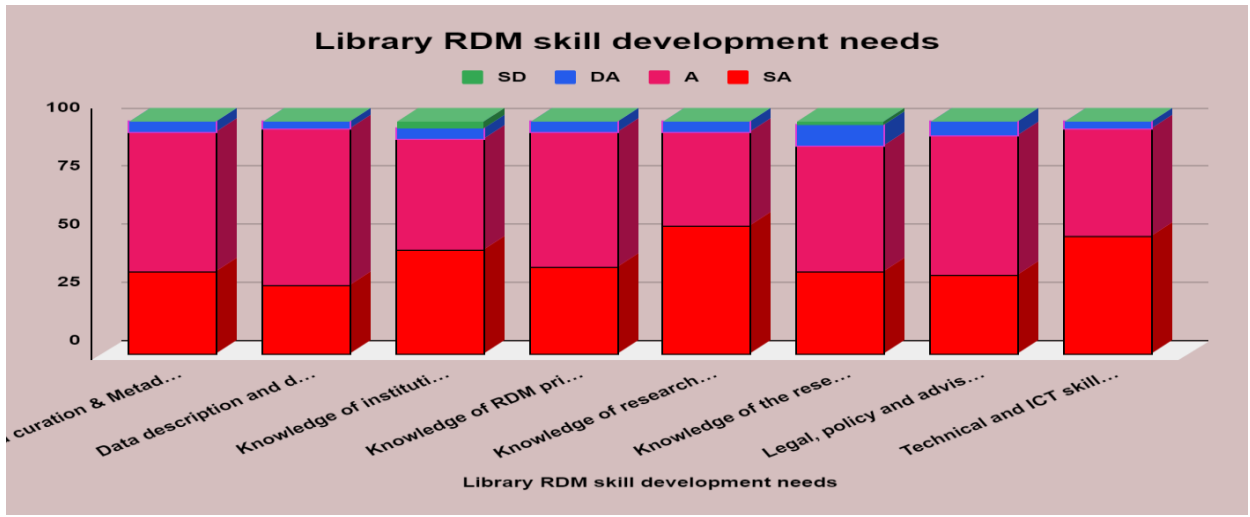

The above table and figure 5 reveal that library professional skill development needs for research data management services. The study analysis shows that most of them 55.2% strongly agreed skill is knowledge of researchers' needs and knowledge of available materials followed by 50.7% technical and ICT skills (e.g., data storage, infrastructure, architecture etc.). Library professionals are mostly agreed skills are data description and documentation 67.2%, followed by data curation and metadata skills 59.7% equally legal, policy and advisory skills (e.g., intellectual property, ethics, licensing etc.), Similarly knowledge of RDM principles and relevant technologies and processes, metadata 58.2%. Almost all of the above-mentioned skills are important based on their responses because only less than 10% disagree and strongly disagree. It shows more than 90% of library professionals agree and strongly agree.

**Table 6. Skills and competencies required for RDM services**

| Skills and competencies required for RDM services | Strongly Agree | Agree | Disagree | Strongly Disagree |
|---|---|---|---|---|
| Big data analytics | 25 (37.3%) | 31 (46.3%) | 10 (14.9%) | 1 (1.5%) |
| Building a repository system | 29 (43.3%) | 34 (50.7%) | 3 (4.5%) | 1 (1.5%) |
| Data ethics | 24 (35.8%) | 38 (56.7%) | 4 (6%) | 1 (1.5%) |
| Data management planning | 26 (38.8%) | 38 (56.7%) | 3 (4.5%) | 0 (0%) |
| Data visualization | 20 (29.8%) | 42 (62.7%) | 5 (7.5%) | 0 (0%) |
| Identifying data repositories for various subject areas | 29 (43.3%) | 34 (50.7%) | 3 (4.5%) | 1 (1.5%) |
| Deep learning techniques | 21 (31.3%) | 34 (50.7%) | 9 (13.5%) | 3 (4.5%) |
| Metadata standards | 25 (37.3%) | 39 (58.2%) | 3 (4.5%) | 0 (0%) |
| Proficiency in qualitative analysis & statistical analysis | 24 (35.8%) | 39 (58.2%) | 4 (6%) | 0 (0%) |
| Understanding of various types of data structure and file formats | 26 (38.8%) | 35 (52.2%) | 5 (7.5%) | 1 (1.5%) |

**Figure 6. Skills and competencies required for RDM services**

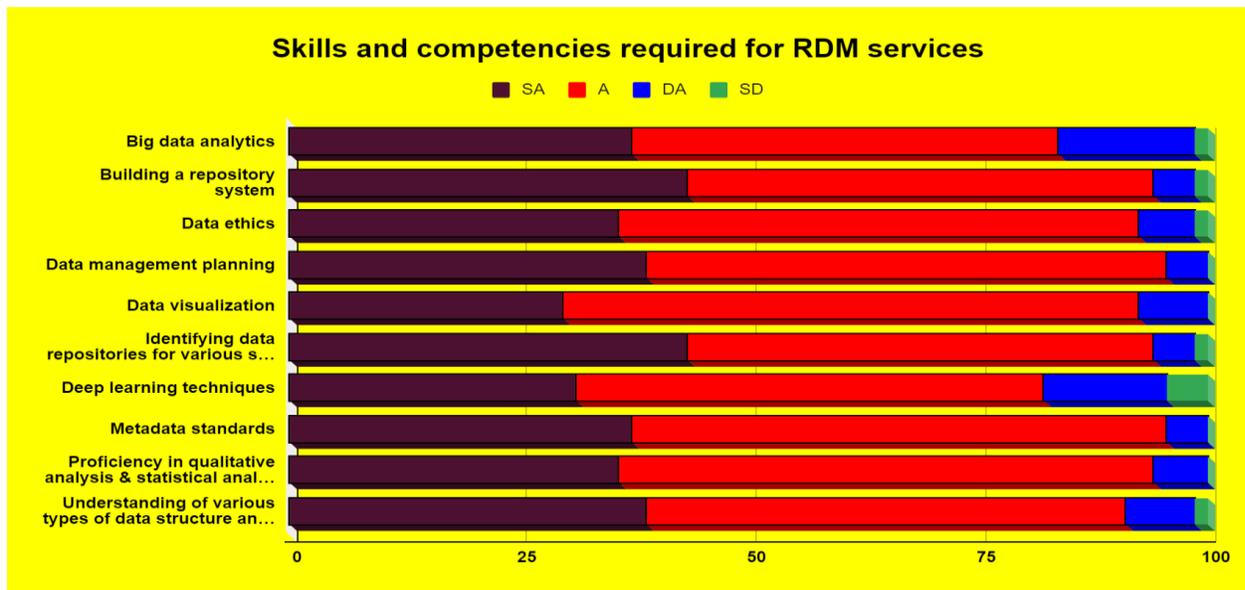

Table and figure 6 discussed skills and competencies required for providing RDM services. The respondents' opinions shows that other than big data analytics remaining all the skills are respondents agreed more than 50% its important the skills are: agreed data visualization (62.7%), metadata standards and proficiency in qualitative analysis and statistical analysis (58.2%), data ethics and data management planning (56.7%), understanding of various types of data structure and file formats (52.2%) building a repository system and deep learning techniques (50.7%). In this result highly recommended skill is data visualization skill.

**Table 7. Approaches to staff RDM skills development in libraries**

| Approaches to staff RDM skills development in libraries | Frequency | Percentage (n=67) |
|---|---|---|
| In-house staff workshops and/or presentations | 43 | 64.2 |
| Support for staff to take courses related to research data services | 46 | 68.7 |
| Support for staff to attend conferences or workshops on research data management | 47 | 70.1 |
| Collaboration with an academic program to develop professionals with skills related to research data services | 38 | 56.7 |

**Figure 7. Approaches to staff RDM skills development in libraries**

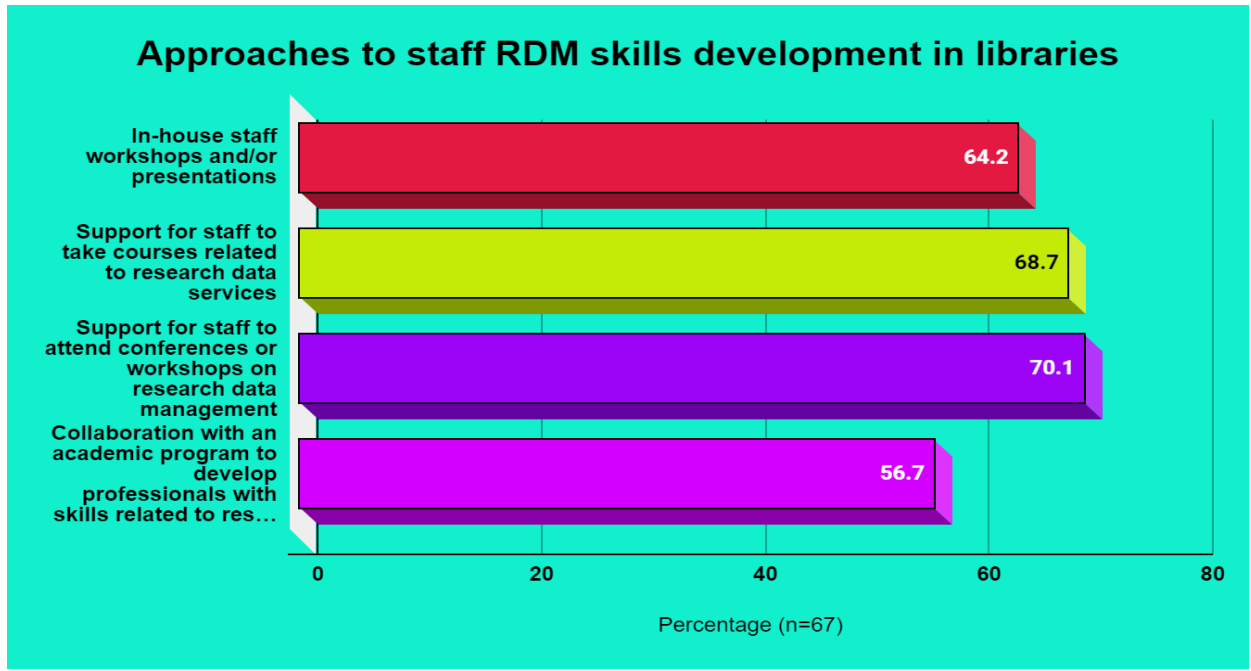

Table and figure 7 narrate on approaches to staff RDM skills development in libraries. The large number of respondents 70.1% expects institutes or funding organizations have to support for staff to attend conferences or workshops on research data management, followed by 68.7% support for staff to take courses related to research data services, 64.2% in-house staff workshops and/or presentations, and 56.7% collaboration with an academic program to develop professionals with skills related to research data services. The overall result reveals they want more tearing and support from the parent body.

Table 8. Managerial Issues and Challenges

| Managerial Issues and Challenges | Strongly Agree | Agree | Disagree | Strongly Disagree |
|---|---|---|---|---|
| Compliance with funder requirements | 14 (20.9%) | 43 (64.2%) | 10 (14.9%) | 0 (0%) |
| Engagement of academic staff | 25 (37.3%) | 39 (58.2%) | 2 (3%) | 1 (1.5%) |
| Infrastructure | 23 (34.3%) | 38 (56.7%) | 6 (9%) | 0 (0%) |
| Lack of budget assigned for research data services | 23 (34.3%) | 31 (46.3%) | 10 (15%) | 3 (4.5%) |
| Lack of technical support for data services | 22 (32.8%) | 32 (47.8%) | 11 (22.4%) | 2 (3%) |
| Lack of top administration & university-level support and prioritization | 22 (32.8%) | 27 (40.3%) | 15 (22.4%) | 3 (4.5%) |
| Legal issues | 13 (19.4%) | 34 (50.8%) | 17 (25.4%) | 3 (4.5%) |
| Limited awareness of data services in academic libraries among "patrons & "librarians" | 20 (29.8%) | 37 (55.2%) | 6 (9%) | 4 (6%) |
| Limited support from liaisons/subject librarians | 22 (32.8%) | 30 (44.8%) | 13 (19.4%) | 2 (3%) |
| Preservation | 17 (25.4%) | 43 (64.2%) | 5 (7.4%) | 2 (3%) |

Figure 8: Managerial Issues and Challenges

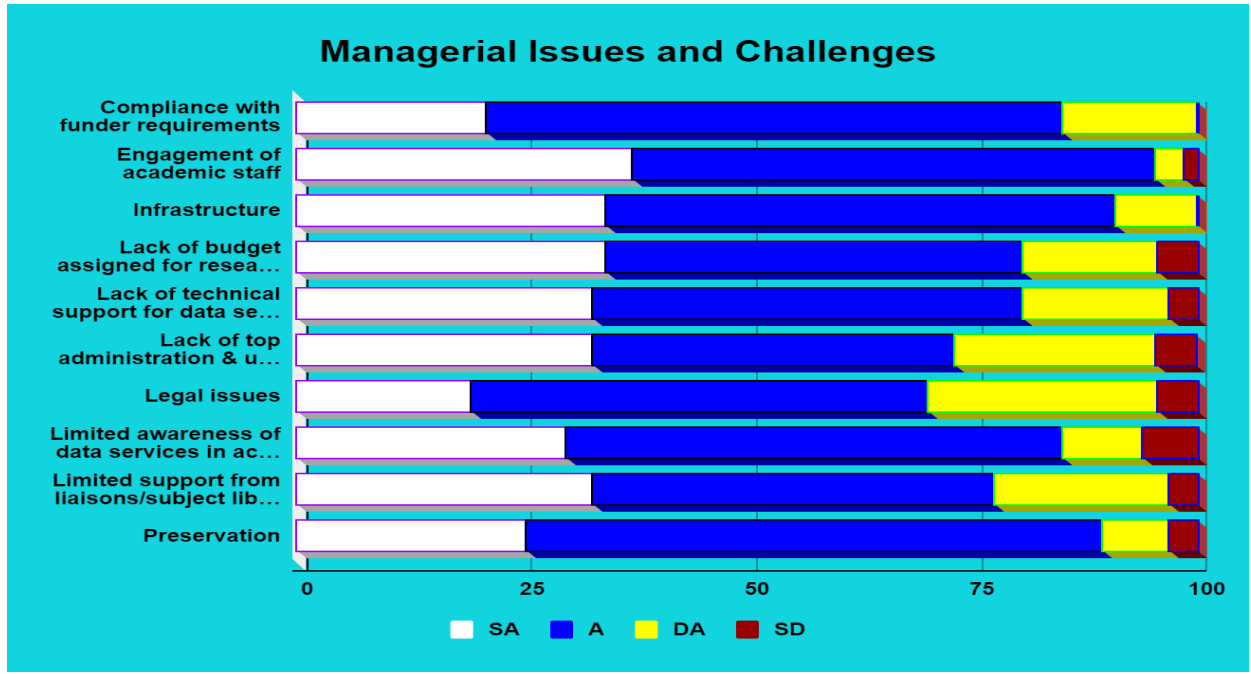

Table and figure 8 summarize the issues and challenges of research data management. The large number of respondents 64.2% agreed compliance with funder requirement and preservation is a major issue as well as challenges. Followed by agreed 58.2% engagement of academic staff, 56.7% infrastructure, 55.2% limited awareness of data services in academic libraries among "patrons & "librarians", 50.8% legal issues, 47.8% lack of technical support for data services, 46.3% lack of budget assigned for research data services, 44.8% limited support from liaisons/subject librarians, 40.3% lack of top administration and university-level support and prioritization.

## 6. Conclusion

The study research data management and related services reported based on the survey result analysis that most of the countries are taking an active part in research data services and providing data management training and/or data literacy instruction and data repositories/ institutional repository kind of services. Whatever ICT development comes for that adequate training is more important to the library professionals before implementing the library. The study accentuates the pre-requisite knowledge of metadata standards, proficiency in qualitative analysis & statistical analysis, and data management planning skill for performing RDM services. To inculcate such skills libraries, support their staff to attend conferences or workshops on research data management or support them to take courses related to research data services. This study lists out various managerial issues like data preservation, Compliance with funder requirements, engagement of academic staff in RDM services, infrastructure, and legal issues, that are most common even though rectification of these issues is equally important to librarians and parenting institutes where they are lacking and do not notice the arising problems, this may affect users and employers of the institute.

Research data is important for the present and future researchers but how it is maintained in a structural way for easy and secure access, and user friendliness these all are important, presents study has divulged major findings and approaches by libraries to perform RDM services and contributed towards future stride taken up by library professionals for the seamless performance of such services.